\documentclass[a4paper,11pt]{article}
\usepackage{pos}

\usepackage{enumitem}   
\usepackage{cancel}

\input{defs}

\definecolor{violet}{rgb}{0.5,0,0.5}

\newcommand{\emdash}{\thinspace---\thinspace}

\title{Effective Lagrangians from functional matching}

\author*[a]{Stefan Dittmaier}
\author[a]{Sebastian Schuhmacher}
\author[a]{Maximilian Stahlhofen}

\affiliation[a]{Albert-Ludwigs-Universit\"at Freiburg,
Physikalisches Institut, \\
Hermann-Herder-Stra\ss{}e 3,
79104 Freiburg, Germany}

\emailAdd{stefan.dittmaier@physik.uni-freiburg.de}
\emailAdd{sebastian.schuhmacher@physik.uni-freiburg.de}
\emailAdd{maximilian.stahlhofen@physik.uni-freiburg.de}

\abstract{We briefly review a variant of functional matching to derive
an Effective Field Theory (EFT) for heavy particles at the one-loop level
in the top-down approach.
The method integrates out heavy fields that correspond to mass eigenstates,
i.e.\ after removing mixing effects by diagonalizing mass matrices.
Tree- and loop-level effects are separated by employing the background-field method,
hard and soft modes are separated with the use of the expansion by regions.
The method is exemplified for the Higgs Singlet Extension of the Standard Model where the
mass~$\MH$ of the additional Higgs boson is considered large, and the Higgs
mixing angle~$\alpha$ is assumed to scale like $1/\MH$, in order to
guarantee decoupling in the large-$\MH$ limit.
Our calculation is agnostic
w.r.t.\ the type (SMEFT vs.\ HEFT) of the emerging EFT.
Eventually the emerging EFT Lagrangian can be transformed into SMEFT form,
but only at the cost of introducing fermionic EFT operators, although no such
operators are directly generated upon solving the functional integral
over the heavy Higgs field.
}

\FullConference{Loops and Legs in Quantum Field Theory (LL2026)\\
12-17, April, 2026\\
Bayreuth, Germany\\}

\begin{document}
\maketitle

\section{Introduction}

Effective Field Theories (EFTs), such as Standard Model (SM) Effective
Field Theory (SMEFT) or Higgs Effective Field Theory (HEFT),
provide an excellent framework to address physics beyond the
Standard Model (BSM) induced by new heavy particles.
In the top-down approach, EFT Lagrangians of underlying UV-complete
models are derived by integrating out the heavy fields, and the
EFT couplings (Wilson coefficients) can be confronted with
constraints from experimental data.

In the two foundational papers~\cite{Dittmaier:2026nnb,Dittmaier:2021fls}
we have proposed a variant of {\it functional matching} where the
heavy fields are integrated out directly in the path integral.
The method, which employs the background-field
method and the expansion by regions, is to some extent a further development
of the method proposed in \citere{Dittmaier:1995cr},
but also shares features with similar proposals in the
literature~\cite{Fuentes-Martin:2016uol,Zhang:2016pja,Cohen:2020fcu}.
Specifically, we have integrated out the heavy Higgs field~H (with mass $\MH$) in a
Higgs Singlet Extension of the SM (SESM) under the assumption
that the Higgs mixing angle~$\alpha$ is suppressed like $1/\MH$
in the limit $\MH\to\infty$, in order to guarantee the decoupling of H in this limit.
Fermions are treated in the massless approximation.
Since the SM-like Higgs doublet
is represented non-linearly, our calculation is agnostic
w.r.t.\ the important question whether the resulting EFT is of
SMEFT or HEFT type. In the the considered large-$\MH$ limit,
we find (and prove) that the SMEFT
framework with only bosonic operators is not sufficient,
although there is no coupling to massless fermions in the BSM sector of the SESM.
The EFT Lagrangian can, however, be brought into SMEFT form upon introducing
fermionic SMEFT operators.
EFTs for singlet-Higgs extensions with heavy Higgs bosons have been discussed in the
literature quite extensively in different variants of the model and the inspected
large-mass limits, with
\citeres{Boggia:2016asg,Buchalla:2016bse,Cohen:2020xca}
close to our setup and
\citeres{Cohen:2020fcu,Ellis:2017jns,Haisch:2020ahr}
at least related; more details on the relation to our work are given in
\citere{Dittmaier:2026nnb}.

In contrast to most discussions of EFT matching procedures, we are very
explicit with respect to the procedure of renormalization, which is
essential when confronting EFT predictions with results of the full
underlying model.
Our calculation is validated by comparing our EFT predictions
with full SESM higher-order predictions for a representative
set of electroweak precision observables.

\section{The Higgs singlet extension of the SM with a heavy Higgs boson}

Before we go over the individual steps of the functional matching method,
we briefly introduce the salient features of the SESM,
entirely following the conventions and notation of
\citeres{Dittmaier:2026nnb,Dittmaier:2021fls},
which are in turn adopted from the formulations of the SM and SESM given in
\citeres{Denner:2019vbn} and \cite{Altenkamp:2018bcs}, respectively.
The SESM extends the SM by a real singlet scalar field
$\sigma=\varv_1+h_1$ with vacuum expectation value (vev) $\varv_1$ and
Higgs excitation $h_1$, which interacts with SM particles via the
portal provided by the Higgs potential
(featuring a $ \mathbb{Z}_2 $ symmetry under $ \sigma\rightarrow-\sigma$)
\begin{align}
V^{\SESM}={}&
- \frac{1}{2}\mu_{2}^2\tr\left[\Phi^\dagger\Phi\right]
+\frac{1}{16}\lambda_{2}  \Big(\tr\left[\Phi^\dagger\Phi\right]\Big)^2
-\mu_{1}^2\sigma^2+\lambda_{1}\sigma^4+\frac{1}{2}\lambda_{12}\sigma^2\tr\left[\Phi^\dagger\Phi\right],
\label{eq:V_SESM}
\end{align}
where $\Phi$ represents the SM-like SU(2) doublet in matrix notation.
We consistently parametrize $\Phi$ non-linearly,
\begin{equation}
\Phi=\frac{1}{\sqrt{2}}\left(\varv_{2}+h_{2}\right)U \,, \qquad
U=\exp\biggl(2\ri\frac{\varphi}{\varv_{2}}\biggr) \,, \qquad
\varphi=\frac{1}{2}\varphi_a\tau_a \,,  \qquad
\tau_a = \mbox{Pauli matrices} \,,
\label{eq:doublet}
\end{equation}
where $\varv_2$ quantifies its vev, $h_2$ is the physical Higgs component, which is a gauge
singlet, and $\varphi_a$ are Goldstone-boson fields.
The two singlet Higgs fields $h_1, h_2$ mix to fields $h$ and $H$ corresponding to mass eigenstates
(with masses $\Mh$ and $\MH$) via a rotation with angle $\alpha$
$ (s_\alpha\equiv\sin\alpha,\,c_\alpha\equiv\cos\alpha) $,
\begin{equation}\label{eq:HfieldRotation}
        \begin{pmatrix}H\\h\end{pmatrix} =
        \begin{pmatrix}c_\alpha&s_\alpha\\-s_\alpha&c_\alpha\end{pmatrix}
        \begin{pmatrix}h_1\\h_2\end{pmatrix}.
\end{equation}
Instead of using the five original parameters $\mu_1^2$, $\mu_2^2$,
$\lambda_1$, $\lambda_2$, and $\lambda_{12}$ of the Higgs potential,
we reparametrize the Higgs sector in terms of parameters that are closer to
phenomenology: the two Higgs masses $\Mh$, $\MH$, the mixing angle $\alpha$,
the vev $\varv_2$, which is closely related to the W-boson mass $\MW$, and
one of the quartic couplings, for which we take $\lambda_{12}$.
To control the large-mass limit, we introduce the parameter $\zeta\equiv\MH/\Mh$
and assume the scaling $\MH\sim\zeta$, $\alpha \sim  s_\alpha \sim\zeta^{-1}$, while $\lambda_{12} \sim 1$
and all SM-like parameters are kept fixed for $\zeta\to\infty$.

\section{Recap of the functional matching method}

\myparagraph{Background-field method (BFM) and non-linear Higgs realization}

Generically, the BFM splits all fields $\phi\to\tilde\phi=\hat\phi+\phi$ into
background parts $\hat\phi$, which are classical (in the sense that they do not appear in loops),
and quantum parts $\phi$,
which correspond to quantum fluctuations and are the integration variables of the path integral.
Diagrammatically, the fields $\hat\phi$ deliver
the tree-like lines and and the fields $\phi$ the lines within loops in Feynman graphs.
Since all vertices that are part of a one-loop subdiagram involve exactly
two quantum fields, at the one-loop level the relevant part of the Lagrangian is
quadratic in quantum fields.

In the non-linear realization of the Higgs doublet $\Phi$, it is convenient
to work with the multiplicative split $U\to\tilde U=\hat U U$ for the
matrix $U$ of the Goldstone-boson fields~\cite{Dittmaier:1995cr}.
Owing to the gauge invariance of the background-field effective action,
it is possible to absorb
all background Goldstone-boson fields into the background
gauge fields by a straightforward St\"uckelberg transformation,
i.e.\ we effectively adopt the unitary background gauge by setting $\hat U\to\bbid$,
which reduces the amount of algebraic work
in the subsequent steps considerably.

\myparagraph{Separation of hard and soft field modes}

Our implementation of the method of regions~\cite{Beneke:1997zp}
for large-mass expansions of Feynman diagrams splits both quantum and
background fields into soft ($s$) and hard ($h$) field modes.
In the SESM with a heavy H~boson, the original field $H$ is split into
$\hat H_s + \hat H_h + H_s + H_h$.
This separation disentangles EFT contributions that correspond to
higher-order effects from hard interactions
ending up in Wilson coefficients of effective operators on one side and effective operators
entering loops built from soft fields on the other.

The splitting of one-loop diagrams into two integration
domains of small and large momenta can be realized as a splitting
of the path integral into two functional integrals extending over
soft and hard field modes,
$\int\mathcal{D}H \to \int\mathcal{D}H_s \int\mathcal{D}H_h$.
While the use of the method of regions was not yet part of
of the procedure of \citere{Dittmaier:1995cr},
where only hard field modes appeared in the calculation of the effective Lagrangian,
it is an integral part of most of the functional methods for integrating out heavy fields
proposed in the
literature~\cite{Fuentes-Martin:2016uol,Zhang:2016pja,Cohen:2020fcu,Dittmaier:2021fls,Dittmaier:2026nnb}
(for more references, see e.g.\ \citere{Dittmaier:2026nnb}).

\myparagraph{Equations of motion (EOMs) for the soft modes of the heavy fields and renormalization}

By design, the EFT is valid for energies where for
any considered process the particle momenta $p_i$ on tree-level lines of Feynman graphs
are much smaller than the heavy masses, i.e.\ $|p^\mu_i|\ll\MH$ in our case.
In other words, background fields only possess soft modes, and
the heavy fields carrying these soft momenta do not represent dynamical degrees
of freedom in the EFT.
The EOMs for the soft modes of the heavy field can be solved 
up to the relevant order in the large-mass expansion, and this solution can be used to
eliminate them from the Lagrangian.

This procedure produces the EFT Lagrangian for the SESM in the large-$\MH$ limit valid at tree level. We obtain
\begin{equation}
\L_{\eff}^\tree=\L_{\SM}^\tree
	-\frac{g_2^2s_{\alpha}^2}{8\varv_2}\tilde{h}(\varv_2+\tilde{h})^2 \bigl(\tilde{C}_\mu^a\bigr)^2
	-\frac{s_{\alpha}^2}{8\varv_2^2}\tilde{h}^2\Box\tilde{h}^2
	+\frac{\Mh^2s_{\alpha}^2}{8\varv_2^3}\tilde{h}^3(6\varv_2^2+7\tilde{h}\varv_2+2\tilde{h}^2)
    \,+\,\mathcal{O}\bigl(\zeta^{-4}\bigr) \,.
\label{eq:LeffTreelong}
\end{equation}
Here, $g_2$ denotes the SU(2) gauge coupling, and the vector fields $\tilde{C}_\mu^a$ are the combination
of SU(2) gauge fields $W_\mu^a$ and the U(1) gauge field $B_\mu$ corresponding to
the massive $\PW^\pm$ and $\PZ$~bosons,
\begin{equation}
	C_\mu=
	\frac{\tau_a}{2}C^a_\mu
	=W_\mu+\frac{\sw}{\cw}B_\mu\frac{\tau_3}{2}
	=\frac{1}{2}\left(W^1_\mu\tau_1+W^2_\mu\tau_2+\frac{1}{\cw}Z_\mu\tau_3\right).
\end{equation}
By virtue of the EOM of the light Higgs field, the tree-level effective Lagrangian
$\L_{\eff}^\tree$ of Eq.~\refeq{eq:LeffTreelong} is equivalent to
$\L_{\eff}^{\tree} =\L_{\SM}^\tree
        -\frac{s_{\alpha}^2}{2\varv_2}\mathcal{O}_{\Phi\Box}^\SMEFT$,
where $\mathcal{O}_{\Phi\Box}^\SMEFT=
\frac{1}{2}\mathrm{tr}\{(\Phi^\dagger\Phi)\Box(\Phi^\dagger\Phi)\}$ is one of the usual
SMEFT operators of the Warsaw basis~\cite{Grzadkowski:2010es}.

The same procedure can be applied to the counterterm Lagrangian of the full SESM.
Since renormalization constants are already one-loop effects, only background fields
are relevant in this context.
This step requires a proper power-counting of all parameters and fields in the heavy-mass limit.
As emphasized in \citere{Dittmaier:2021fls},
in order to obtain a consistent effective Lagrangian the
large-mass expansion must be carefully performed taking into account that the full-theory renormalization constants
may have a different scaling behaviour
than the corresponding renormalized quantities.
``Good schemes'' should respect the decoupling behaviour of the theory at
leading mass order also at higher orders in the EFT expansion.
In this context, the choice of a proper
renormalization of the Higgs vevs, i.e.\ a proper tadpole scheme, is crucial in $\MSbar$
renormalization of full-theory parameters~\cite{Dittmaier:2021fls}.

\myparagraph{Integrating out the hard modes of the heavy quantum fields in the path integral}

Next, we deal with the hard modes of the heavy quantum fields.
Since the part of the Lagrangian that is relevant at one-loop order
is quadratic in the quantum fields, the path integral over the
hard field modes of the
heavy quantum fields is of Gaussian type and can be done analytically.
The major complication in this step is the fact that there are
also terms that are linear in the heavy quantum fields.
These terms can be removed by a field redefinition of the hard quantum field modes
in a fully algorithmic manner.
This means that the resulting part of the Lagrangian quadratic in the hard modes of the heavy quantum fields
can be directly identified based again on a simple power-counting argument
and the large-mass expansion can be directly carried out via a Neumann
series~\cite{Fuentes-Martin:2016uol,Zhang:2016pja,Cohen:2020fcu,Dittmaier:2021fls,Dittmaier:2026nnb},
which establishes another technical improvement
over the procedure described in \citere{Dittmaier:1995cr}.

Schematically, integrating out the hard quantum field~$H_h$ and performing the large-mass
expansion, turns the Lagrangian $\mathcal{L}^\oneloop(\hat\phi,\phi, H_h)$ that is bilinear in
quantum fields into the
effective Lagrangian $\mathcal{L}^\oneloop_\eff(\hat\phi,\phi)$
\begin{align}
\int\mathcal{D}\phi\int\mathcal{D}H_h\,\exp\biggl\{\ri\int\rd^D x\,
\mathcal{L}^\oneloop(\hat\phi,\phi,H_h)\biggr\}
\to
\int\mathcal{D}\phi\,\exp\biggl\{\ri\int\rd^D x\,
\mathcal{L}^\oneloop_\eff(\hat\phi,\phi)\biggr\} \,.
\end{align}
Note that (after the field redefinition) the functional integration over the hard modes of the light quantum fields ($\phi_h$) does not contribute to $\mathcal{L}^\oneloop_\eff$ in dimensional
regularization, because those corrections are related to scaleless one-loop diagrams.

\myparagraph{Final form of the effective Lagrangian}

The effective Lagrangian resulting from the previous steps
only involves soft background and quantum fields,
but none of the modes of the heavy fields.
The effects of the latter are absorbed into the Wilson coefficients
of the effective operators.

Finally, redefinitions of the light quantum fields and IBP in the effective Lagrangian may be used to eliminate redundant operators that only influence
off-shell Green functions, but no physical scattering amplitudes. This step
is, in particular, necessary to bring the effective Lagrangian into canonical form,
i.e.\ to SMEFT  or (more generally) HEFT form.

\section{Structure and validation of the resulting EFT Lagrangian for the SESM}

In \citere{Dittmaier:2026nnb} we have given two forms for the resulting EFT Lagrangian:%
\footnote{In the first preprint version of \citere{Dittmaier:2026nnb} only the first
form had been given. The possibility to transform it to SMEFT form was pointed out to us
by Gerhard Buchalla, who is gratefully acknowledged for this.}
the first form contains only bosonic operators apart from the SM parts, but necessarily
includes EFT operators of non-SMEFT form;
the second form contains only SMEFT operators, but involves also fermionic operators
that are not directly generated by integrating out the heavy~$H$ field.
The fermionic SMEFT operators in the latter case result from the EOMs of the gauge-boson fields,
which mix bosonic and fermionic operators.
From this formal perspective the occurrence of fermionic SMEFT operators is not
surprising.
Nevertheless, we believe it is worth emphasizing in the context of data analyses within
SMEFT that BSM effects may show up in fermionic operators even if the BSM sector does not
directly couple to fermions,  as it is the case in the SESM with massless fermions.

Form~1 of the one-loop EFT Lagrangian is given in the form
\begin{align}
\delta\L_\eff^{\oneloop,\BSM} =
\sum_{i\in \text{bos.SMEFT}} C_i^{\SMEFT} \mathcal{O}_i^{\SMEFT}
+ \sum_{n=1}^8 C_n^{\nonSMEFT} \mathcal{O}_n^{\nonSMEFT},
\end{align}
where the sum in the first SMEFT-like part runs over the bosonic SMEFT operators
$\mathcal{O}^{\SMEFT}_{\Phi}$, $\mathcal{O}^{\SMEFT}_{\Phi\Box}$,
$\mathcal{O}^{\SMEFT}_{\Phi D}$, $\mathcal{O}^{\SMEFT}_{\Phi W}$,
$\mathcal{O}^{\SMEFT}_{\Phi B}$, $\mathcal{O}^{\SMEFT}_{\Phi WB}$
as defined in Sec.~2.4 of \citere{Denner:2019vbn}.
The corresponding Wilson coefficients look like
\begin{align}
C_{\Phi WB}^{\SMEFT}={}&
-\frac{e^2s_\alpha^2I_{20}}{16\pi^2 D(D-2)\sw\cw \varv_2^2}\,, \quad
I_{20} = \frac{1}{\eps}-\gamma_\mathrm{E}+\ln(4\pi)
+\ln\left(\frac{\mu^2}{\MH^2}\right)+\mathcal{O}(\eps) \,,
\end{align}
with the two-point vacuum integral $I_{20}\equiv B_0(0,\MH^2,\MH^2)$
in $D=4-2\eps$ space-time dimensions, the electromagnetic charge $e$, $\sw$ and $\cw$ denoting sine and cosine of the weak mixing angle, respectively, and the unphysical (matching) scale $\mu$ in dimensional regularization.
Only the Wilson coefficient $C_{\Phi\Box}^{\SMEFT}$ involves a contribution
from BSM renormalization constants (specifically from $\delta s_\alpha$),
because only the EFT operator $\mathcal{O}_{\Phi\Box}^{\SMEFT}$ appears
at tree level.
As renormalization scheme for $s_\alpha$, we admit any scheme in which the renormalization constant
$\delta s_\alpha$ shows the same scaling in $\zeta$ as $s_\alpha$ at lowest order\emdash a requirement that is, e.g., met by the on-shell scheme suggested in
\citere{Denner:2018opp}.%
\footnote{In the on-shell scheme used in Ref.~\cite{Dittmaier:2021fls} for $\sa$ the renormalization constant $\delta\sa$ 
contains hard and soft contributions and
depends on the tadpole scheme. This entails differences whether the EFT
calculation of observables is carried out in the linear or non-linear Higgs realization.
Since we have employed the \textit{Parameter Renormalized Tadpole Scheme (PRTS)} 
(see, e.g., Ref.~\cite{Denner:2019vbn})
in the EFT construction in the non-linear realization,
any direct application of our EFT using the linear Higgs realization implies the {\it Gauge-Invariant Vacuum expection value Scheme 
(GIVS)}~\cite{Dittmaier:2022maf}
for the tadpole treatment if $\delta\sa$ is kept unchanged (see 
Ref.~\cite{Dittmaier:2026nnb} for further details).}
In unitary gauge, the non-SMEFT operators exemplarily read
\begin{align}
  \mathcal{O}^{\nonSMEFT}_1
={}& (\varv_2+h)^2(D_B^\mu C_\mu)^a(D_B^\nu C_\nu)^a\, , \quad
(D_B^\mu C_\nu)^a=\partial^\mu C_\nu^a+g_1\eps^{3ab}B^\mu C_\nu^b \,,
\quad \dots \nn\\
  \mathcal{O}^{\nonSMEFT}_8
={}& (\varv_2+h)^2\left[(C_\mu^a)^2(C_\nu^b)^2-C_\mu^aC^{\mu,b}C_\nu^aC^{\nu,b}\right].
\end{align}
The corresponding Wilson coefficients look very much like
$C_{\Phi WB}^{\SMEFT}$ given above.
While the SMEFT Wilson coefficients $C_i^{\SMEFT}$ do not only contain the
suppression factor $s_\alpha^2$, but also factors like $\lambda_{12}^2\varv_2^2/\MH^2$,
the non-SMEFT coefficients $C_n^{\nonSMEFT}$ are all proportional to $s_\alpha^2$,
showing that this part is entirely due to mixing effects.

Form~2 of the one-loop EFT Lagrangian reads
\begin{align}
\delta\L_\eff^{\oneloop,\BSM} =
\sum_{i\in \text{SMEFT}} \hat C_i \,\mathcal{O}_i^\SMEFT\,,
\label{eq:SMEFTform}
\end{align}
where the sum now also includes the fermionic SMEFT operators
$\mathcal{O}^{(1), \SMEFT}_{\Phi F}$,
$\mathcal{O}^{(3), \SMEFT}_{\Phi F}$, and
$\mathcal{O}_{\Phi f}^\SMEFT$ with left-handed fermion doublets $F=L,Q$ and
right-handed fermion singlets $f=l,u,d$~\cite{Denner:2019vbn}.
Four-fermion operators do not occur.
The Wilson coefficients in Eq.~\eqref{eq:SMEFTform} read
\begin{align}
  \hat C_\Phi ={}&
  \frac{(D^2+6D+20) \Mh^4 \sa^2 I_{20}}{2D(D^2-4) \pi^2 \varv_2^6}
  - \frac{(D^2+4) e^2 \Mh^2 \sa^2 I_{20}}{4D(D^2-4) \pi^2 \sw^2 \varv_2^4}
  + \frac{(D-4) (\MH^2 \sa^2 - 2 \lambda_{12} \varv_2^2)^3 I_{20}}{48 \MH^2 \pi^2 \varv_2^6}
  \nn\\ & {}
  + \frac{(D-4)(D+2) \Mh^2 \sa^2 (\MH^2 \sa^2 - 2 \lambda_{12} \varv_2^2)
    I_{20}}{4D(D-2) \pi^2 \varv_2^6} \,,
  \nn\\
  \hat C_{\Phi\Box} ={}& -\frac{s_\alpha\delta s_\alpha}{\varv_2^2}
  - \frac{(D^2+4) e^2 \sa^2(3-2\sw^2) I_{20}}{16D(D^2-4) \pi^2 \sw^2 \cw^2 \varv_2^2}
  + \frac{(\MH^2 \sa^2 - 2 \lambda_{12} \varv_2^2) I_{20}}{192D(D-2)\MH^2 \pi^2 \varv_2^4}
  \nn\\ & \qquad {}
  \times \left[(D^3-18D^2+128D-96) \MH^2 \sa^2
  - 2D(D^2-42D+44) \lambda_{12} \varv_2^2\right] ,
  \nn\\
  \hat C_{\Phi D} ={}&
  -\frac{(D^2+4) e^2 \sa^2 I_{20}}{4D(D^2-4) \pi^2 \cw^2 \varv_2^2} \,,
  \qquad
  \hat C_{\Phi W} = -\frac{(D-4) e^2 \sa^2 I_{20}}{8D (D^2-4) \pi^2 \sw^2 \varv_2^2} \,,
  \nn\\
  \hat C_{\Phi B} ={}&  \frac{\sw^2}{\cw^2} \hat C_{\Phi W} \,,
  \qquad
  \hat C_{\Phi WB} = \frac{2\sw}{\cw} \hat C_{\Phi W}\,,
  \qquad
  \hat C_{W} = 0 \,,
  \nn\\
  \hat C^{(3)}_{\Phi F} ={}& \frac{(D-6)e^2\sa^2 I_{20}}{16D(D^2-4)\pi^2\sw^2 \varv_2^2} \,,
\qquad
  \hat C^{(1)}_{\Phi F} = \frac{\sw^2}{\cw^2} \hat C^{(3)}_{\Phi F} \, Y_{\rw,F}
  \,,
\qquad
  \hat C_{\Phi f} = \frac{\sw^2}{\cw^2} \hat C^{(3)}_{\Phi F} \, Y_{\rw,f} \,.
  \label{eq:Chat}
\end{align}
Note that the Wilson coefficients $\hat C_i$ of the bosonic operators
do not exactly coincide with the $C_i^\SMEFT$ of the first form of the EFT Lagrangian,
but their structure is very similar.
All non-zero fermionic Wilson coefficients are proportional to $\sa^2$,
and the ones of $\mathcal{O}^{(1),\SMEFT}_{\Phi F}$ and $\mathcal{O}_{\Phi f}^\SMEFT$
receive factors of the weak hypercharges $Y_{F/f}$ of the respective fermions.

\begin{figure}
        \centering{\includegraphics[width=.9\textwidth]{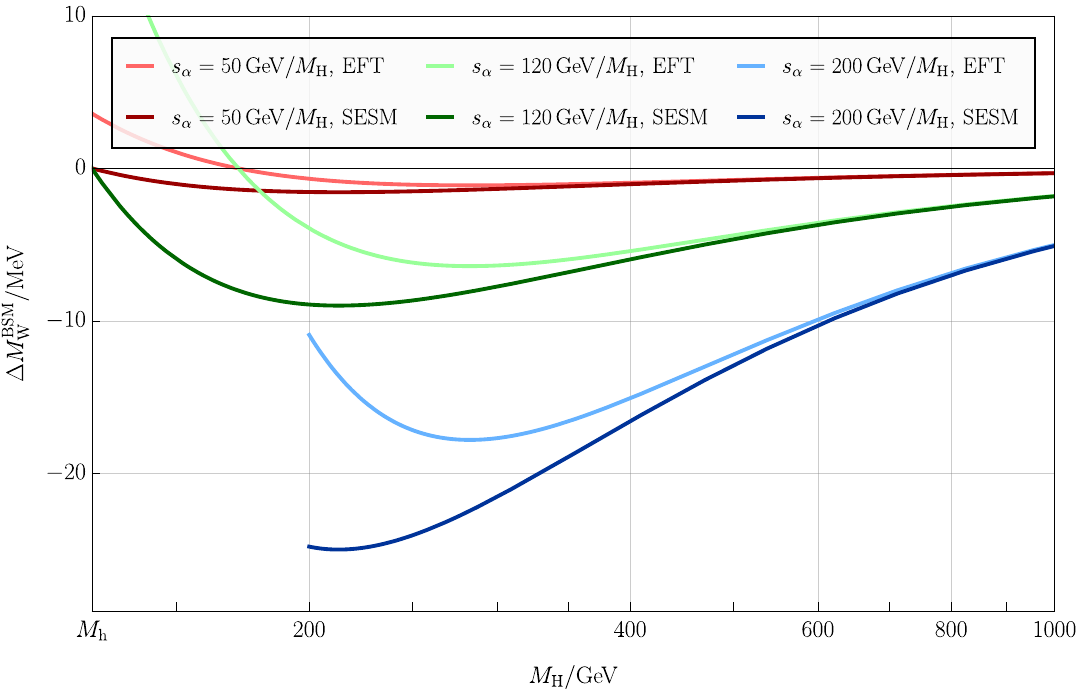}}
        \caption{BSM correction $\Delta\MW^{\BSM}$ to the W-boson mass,
shown for the full SESM and the EFT approximation.
The product $s_\alpha\MH$ is kept fixed in each curve.
(Taken from \citere{Dittmaier:2026nnb}.)}
\label{fig:MW}
\end{figure}

We have validated both forms of
the effective Lagrangian at next-to-leading order
in the coupling expansion by verifying that
the difference between EFT and full-theory predictions
for several observables vanishes faster than $1/\MH^2$ in the large-$\MH$ limit.
Specifically, we examined the
W-boson mass $\MW$ derived from muon decay, some W/Z-boson decay widths, the effective weak mixing angle,
and the leptonic four-body Higgs decay $\Ph\to\PW\PW\to\nu_\Pe\Pe^+\mu^-\overline{\nu}_\mu$ 
in \citere{Dittmaier:2026nnb}.
In \reffi{fig:MW} we exemplarily show the BSM contribution to $\MW$ to illustrate this fact.

\section{Conclusions}

We briefly summarize the salient features and
particular strengths of our method~\cite{Dittmaier:2021fls,Dittmaier:2026nnb} to derive the effective Lagrangian
at the one-loop level, some of them common to the approaches
of \citeres{Fuentes-Martin:2016uol,Zhang:2016pja,Cohen:2020fcu}:
\begin{myenumerate}
	\renewcommand{\labelenumi}{\theenumi}
	\renewcommand{\theenumi}{(\roman{enumi})}
	\item
	a clear separation of tree-level and loop effects of the heavy field modes by employing the BFM;
	\item
	the possibility to fix the (background) gauge in intermediate steps of the
	calculation and to restore gauge invariance of the effective Lagrangian at the end;
	\item
	transparency in the sense that at each stage of the calculation it is possible to
	identify the origin of all contributions to the effective Lagrangian in terms of (classes of)
	Feynman diagrams;
	\item
	flexibility due to the fact that no ansatz is made for the effective Lagrangian.
	Whether the emerging EFT is of SMEFT or HEFT type is part of the result.
	\item
	An automation of the method is possible, since it is fully algorithmic. In principle,
	given a BSM Lagrangian, a proper definition of the large-mass limit with a corresponding power-counting scheme,
	and some details on the renormalization
	of the BSM sector, the actual determination of the effective Lagrangian
	at the one-loop level can be carried out by computer algebra.
\end{myenumerate}
As an illustrative example we have derived the EFT Lagrangian to order ${\cal O}(1/\MH^2)$
for the SESM with massless fermions and
a heavy Higgs boson of mass $\MH$ in the ``weak-coupling limit''
where the Higgs mixing angle $\alpha$ is suppressed with $1/\MH$.
We have written the emerging EFT Lagrangian in two alternative forms:
one which is of SMEFT form, but involves both bosonic and fermionic SMEFT operators,
and another one that involves only bosonic EFT operators, albeit not all
of SMEFT type.

\setlength{\bibsep}{2.0pt}

\end{document}